\documentclass[conference]{IEEEtran}
\usepackage[style=ieee]{biblatex}
\usepackage{amsmath,amssymb,amsfonts}
\usepackage{algorithmic}
\usepackage{graphicx}
\usepackage{textcomp}
\usepackage{xcolor}

\def\BibTeX{{\rm B\kern-.05em{\sc i\kern-.025em b}\kern-.08em
    T\kern-.1667em\lower.7ex\hbox{E}\kern-.125emX}}

\bibliography{references.bib}

\begin{document}

\title{A Review on Message Complexity of the Algorithms for Clock Synchronization in Distributed Systems
}

\author{
    \IEEEauthorblockN{Chandeepa Dissanayake\IEEEauthorrefmark{1}, Chanuka Algama\IEEEauthorrefmark{2}}
    \IEEEauthorblockA{
        \textit{Department of Statistics and Computer Science} \\
        \textit{University of Kelaniya}\\
        Dalugama, Sri Lanka \\
        \IEEEauthorrefmark{1}chandeepadissanayake@gmail.com, \IEEEauthorrefmark{2}chanukaravishan25@gmail.com
    }
}

\maketitle

\begin{abstract}
In this work, we present an extensive analysis of clock synchronization algorithms, with a specific focus on message complexity. We begin by introducing fundamental concepts in clock synchronization, such as the Byzantine generals problem and specific concepts like clock accuracy, precision, skew, offset, timestamping, and clock drift estimation. Describing the concept of logical clocks, their implementation in distributed systems is discussed, highlighting their significance and various approaches. The paper then examines four prominent clock synchronization algorithms: Lamport's Algorithm, Ricart–Agrawala Algorithm, Vector Clocks Algorithm, and Christian's Algorithm. Special attention is given to the analysis of message complexity, providing insights into the efficiency of each algorithm. Finally, we compare the message complexities of the discussed algorithms.
\end{abstract}


\section{Introduction}
Clock synchronization in distributed systems refers to the process of ensuring that the clocks of multiple computers or devices in a network are closely aligned, despite differences in their individual clock speeds and delays. Due to the factors such as the clock drift, hardware variations, and network delays, inherent inaccuracies are unavoidable in these systems, which in turn emphasized the need for clock synchronization. In the absence of clock synchronization, different time based (and possibly date based) fields such as timestamps become inconsistent and unreliable.\par

Message complexity refers to the total number of messages exchanged in a distributed system, such as, and most importantly, during the execution of a synchronization algorithm. It's a critical factor in evaluating the efficiency of clock synchronization algorithms.\par

However, minimizing message complexity must be balanced with achieving accurate synchronization. Some algorithms may require more messages to achieve higher accuracy or to handle more complex scenarios, such as systems with high clock drift or systems that require fault tolerance\cite{Ghosh2012}.

In clock synchronization, a low message complexity is preferable because it reduces network traffic and minimizes the resources utilized for communication. A high message complexity can lead to network congestion, increased latency, and reduced overall system performance.\par

Accordingly, it is evident that clock synchronization is a topic of interest and a concern of many academics as well as in the industrial aspects, not only for the distributed systems but for the entire discipline of computer science. In this regard, understanding message complexity is crucial in the design and selection of clock synchronization algorithms, as it directly impacts network traffic, system performance, and synchronization accuracy.

\section{Background of the Clock Synchronization Problem \& Fundamental Concepts}
Clock synchronization is an important and fundamental concept in distributed systems as it allows for the coordination of processes across different nodes. In a distributed system, each node operates its own internal clock. Over time, these clocks can drift apart due to variations in their counting rates, leading to a phenomenon known as clock drift. This can cause significant issues within the system, particularly regarding the correct ordering of events and accurate timestamping\cite{Lamport1978}.\par

For example, in a distributed computing environment, accurate global time is necessary for the system to function efficiently. Erroneous outcomes can occur if the clocks are not synchronized, especially in processes that rely heavily on precise timing, such as transaction processing or data synchronization\cite{Mills1991}.\par

To illustrate, consider a Unix system that uses the make command to compile new or modified code (not the only kernel that support make program). The make command depends on the system clock to determine the source files that are needed to be recompiled. If the source files are on a separate file server and the two machines have  clocks that are not synchronized, the make program might not produce the correct results\cite{Stallings2014}. Depending on the purview of the reader, this might be portrayed as one of the most simplest and unconvincing examples that someone could ever produce, but it should be sufficient to emphasize the extent of catastrophes it could result in the absence of a proper synchronization.\par

In summary, clock synchronization is vital for maintaining order and accuracy in systems where multiple independent nodes are operating simultaneously. It ensures that all nodes in a distributed system can cooperate effectively, thereby enhancing the overall performance and reliability of the system \cite{Kopetz1997}.

\subsection{The Byzantine Generals Problem}
    The Byzantine Generals Problem (BGP) is an influential problem in the field of distributed computing that bears a close resemblance to the fundamental need for synchronization, including clock synchronization\cite{Lamport1982}. Both problems concern the coordination of independent agents in a system, be they clocks in a network or generals in an army. In BGP, the generals must agree on a common action, akin to how clocks in a network must agree on a common time. However, the problems differ in the type of agreement sought and the challenges faced in achieving this agreement.\par

\begin{figure*}[!t]
        \centering
        \includegraphics[width=0.75\textwidth]{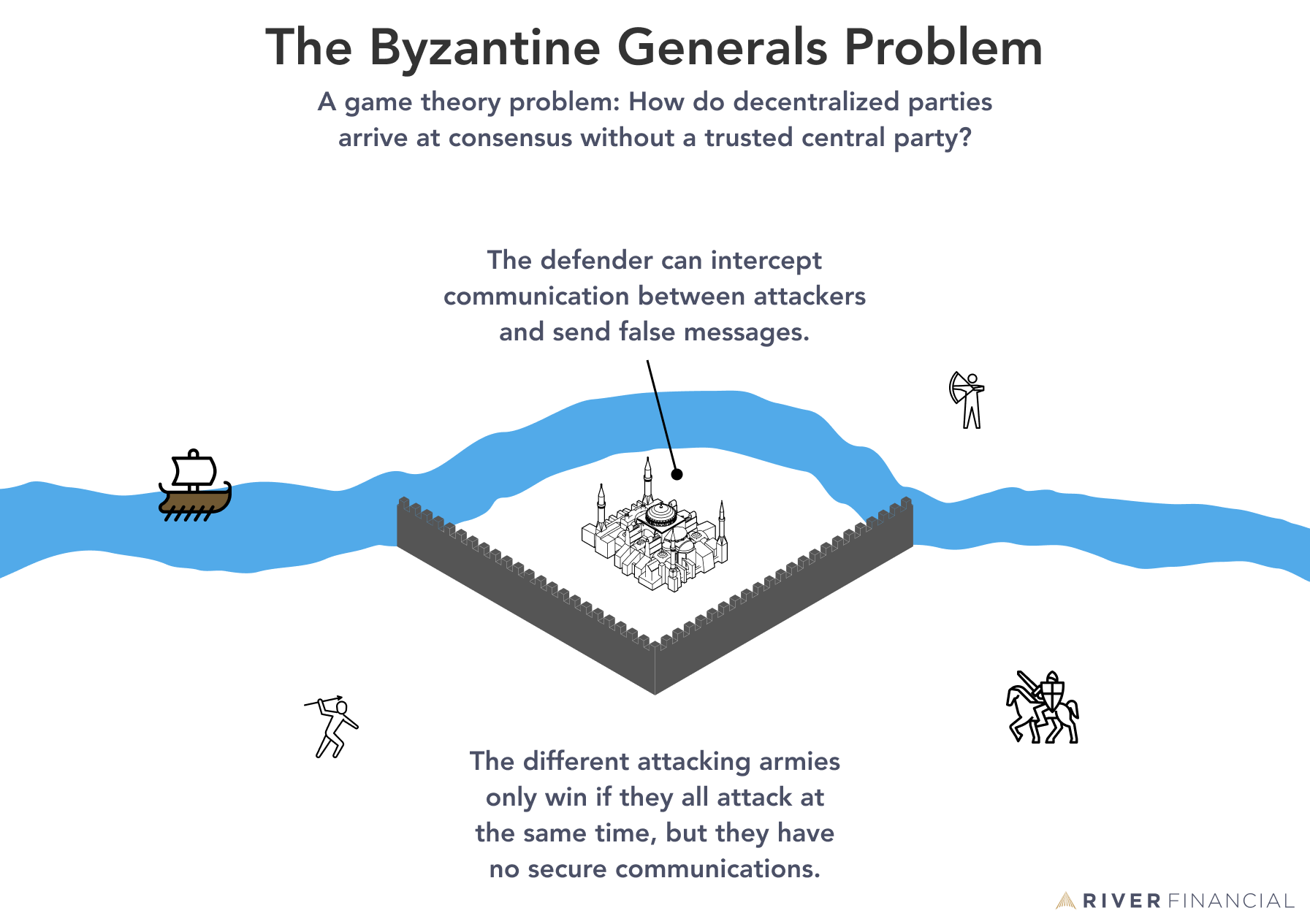}
        \caption{Figure: Byzantine Generals Problem\cite{river}}
        \label{fig:diagram_byzantine_1}
    \end{figure*}

    In the Byzantine Generals Problem illustrated in \ref{fig:diagram_byzantine_1}, a group of generals, each commanding a division of the Byzantine army, encircle a city. They can only conquer the city if they all attack at the same time. However, they can only communicate with each other by sending messengers, and some of these generals may be traitors who send false messages. The challenge, then, is for the loyal generals to reach a consensus on whether to attack or retreat, despite the presence of traitors\cite{Lamport1982}.\par

    The Byzantine Generals Problem sparked the need for clock synchronization in computing because it highlighted the challenges inherent in achieving consensus in a distributed system. Clock synchronization is a form of consensus problem: all nodes in a system must agree on the time. If some nodes provide incorrect time information—either due to faults, delays, or malicious intent—it can lead to similar issues as traitorous generals in the BGP. Therefore, solutions to the BGP, such as Byzantine fault tolerance algorithms, also inform strategies for clock synchronization\cite{Castro1999, Lamport1982}.

\subsection{Clock Accuracy \& Clock Precision}
    Clocks in computer systems are typically based on quartz oscillators or atomic clocks. Clock accuracy refers to how closely a clock measures true time, while precision describes how consistent are clock readings, over time. Both accuracy and precision play significant roles in synchronization\cite{Kopetz1997, Stallings2014}.\par

    Quartz clocks employ an electronic oscillator that is controlled by a quartz crystal to maintain accurate timekeeping. The oscillation of the quartz crystal generates a signal with an extremely precise frequency, resulting in quartz clocks being notably more precise compared to traditional and mechanical clocks.\cite{Ashby2003}.\par

    Meanwhile, atomic clocks utilize atomic vibrations to measure time. These clocks are even more accurate than quartz clocks due to the factors such as atomic resonance's high quality factor, which is the ratio of the resonance's absolute frequency to the line-width of the resonance itself. Atomic clocks are also less sensitive to environmental effects such as temperature changes, further enhancing their accuracy\cite{Lombardi2001}.\par

    Both the accuracy and precision of a clock are crucial for synchronization in distributed systems. A highly accurate clock ensures that the measured time closely matches the true time, while a precise clock ensures that the time readings are consistent over time. If a clock's accuracy or precision is compromised, it can lead to synchronization errors that disrupt the functioning of the distributed system \cite{Kopetz1997, Stallings2014}.

\subsection{Clock Skew and Offset}
    Clock skew refers to the relative difference in the rates at which individual clocks advance, while clock offset represents the difference in absolute time values between clocks\cite{Kopetz1997, Dwork1988}. These factors introduce significant challenges in achieving synchronization in distributed systems.

    The clock skew and clock offset can be more formally represented as follows:
    Given two clocks, $C_1$ and $C_2$, with time readings $t_1$ and $t_2$ respectively, and T representing the true time:
    \begin{enumerate}
        \item Clock Skew: If $T_{C_1}$ and $T_{C_2}$ represent the times at which a clock signal arrives at $C_1$ and $C_2$ respectively, the skew $S$ between the two clocks can be defined as:
            \begin{equation}
                S = T_{C_1} - T_{C_2}
            \end{equation}
        \item Clock Offset: The offset D between the two clocks can be defined as:
            \begin{equation}
                D = t_1 - t_2
            \end{equation}
    \end{enumerate}\par

    Clock skew, or the difference in clock rates, can be caused by a variety of factors, including temperature variations, voltage changes, and manufacturing differences in the clock circuits. Over time, this skew can cause the clocks in a distributed system to drift apart, leading to synchronization problems. For instance, if two processes are supposed to run simultaneously based on their local clocks, a skew between these clocks could cause one process to execute before the other, potentially leading to incorrect results\cite{Sundararaj2005}.\par

    Effective clock synchronization protocols need to account for both clock skew and offset. By compensating for these factors, these protocols can help ensure that all clocks in a distributed system maintain a consistent time, thereby improving the system's overall performance and reliability \cite{Ghosh2014}.

\subsection{Time Synchronization Protocols}
    Various protocols have been developed to synchronize clocks in distributed systems. The Network Time Protocol (NTP) and the Precision Time Protocol (PTP) are widely used in networked environments. These protocols utilize synchronization algorithms and mechanisms to adjust clock rates and minimize clock discrepancies\cite{Mills1991, IEEE1588}.\par

    The Network Time Protocol (NTP) is an Internet protocol used to synchronize computer clock times in a network of computers. NTP uses a hierarchical, semi-layered system of time sources, which are organized in stratum levels. The stratum level defines the distance from the reference clock (source of time). NTP servers at the top of the hierarchy (stratum 1) are synchronized with an atomic clock or GPS\cite{Mills1991}.\par

    The Precision Time Protocol (PTP), that is defined in the IEEE 1588 standard, is designed for local area networks (LANs) that require precise time synchronization. PTP can achieve clock accuracy in the sub-microsecond range, which makes it suitable for control and measurement systems and critical systems where high precision is required\cite{IEEE1588}.\par

    In non-networked environments, or in systems where a network-based time source is not available or reliable, other synchronization protocols can be used. One such protocol is the Reference-Broadcast Synchronization (RBS) protocol. RBS is designed for wireless sensor networks where the communication medium is shared. In RBS, a reference node sends a broadcast message to all other nodes, and these nodes use the arrival time of this message to synchronize their clocks\cite{Elson2002}.\par

    Another such protocol is the Timing-sync Protocol for Sensor Networks (TPSN), which organizes the network into a hierarchical tree structure where the root node has the reference clock. The root node then synchronizes the clock with its children, and this process continues to the depth of the tree until all nodes are synchronized\cite{Ganeriwal2003}.

\subsection{Timestamping and Clock Drift Estimation}
    Timestamping is a fundamental technique in clock synchronization, where events or data packets are marked with timestamps. By comparing these timestamps, clock drift (deviation in clock rates) can be estimated, and adjustments can be made to synchronize clocks\cite{Mills1991}.\par

    Timestamping marks each event or data packet, with the time at which it occurred or was transmitted. This timestamp, which is based on the clock of the system where the event took place or the packet was sent, provides a reference point that can be used to order events or measure the time interval between them. This is crucial in distributed systems, as it allows for the coordination of activities across different nodes in the system. It's also fundamental in network communication for ensuring the correct ordering and timing of data packets\cite{Mills1991, Elson2002}.\par

    Clock drift estimation, on the other hand, is the process of determining the rate at which a clock deviates from a reference time source. In an ideal scenario in a perfect world, all clocks in a distributed system would run at the same rate. However, due to differences in hardware, environmental conditions, and other factors, clocks can drift apart over time, leading to discrepancies in their time readings. By comparing timestamps from different clocks, the rate of this drift can be estimated. Once the drift is known, it can be compensated for, allowing the clocks to be synchronized\cite{Ganeriwal2003, Source3}.\par

    It is worthwhile to note that, a significant amount of research has been carried out on timestamping and clock drift estimation. For instance, \cite{Source3} discusses the precision and stability of different computer system clocks and the typical network delay characteristics. Their work emphasizes the importance of accurate clock drift estimation in achieving precise time synchronization across a distributed system\cite{Source3}.

\subsection{Synchronization Algorithms}
    Synchronization algorithms for distributed systems are crucial to coordinate and integrate the actions of multiple nodes in the system. These algorithms can be broadly classified into two categories: mutual exclusion algorithms and deadlock detection algorithms\cite{Tanenbaum2007}.\par

    Mutual exclusion algorithms ensure that only one process at a time can execute a critical section of code. They prevent race conditions and ensure the consistency and correctness of the system. Some examples of these algorithms include the Ricart-Agrawala algorithm and the Lamport's Bakery algorithm \cite{Ghosh2012}.\par

    Deadlock detection algorithms, on the other hand, help in identifying and resolving deadlocks in the system. Deadlocks can occur when multiple processes are waiting for resources held by each other, leading to a standstill. Some examples of deadlock detection algorithms include the Chandy-Misra-Haas algorithm and the Ho-Ramamoorthy algorithm\cite{Tanenbaum2007}.\par

    These synchronization algorithms are used in a variety of applications, such as database systems, operating systems, and real-time systems, where the correct order of operations is crucial for system performance and reliability \cite{Ghosh2012}.

\subsection{Message Complexity of Clock Synchronization Algorithms}
    Message complexity is a critical factor in evaluating the efficiency of clock synchronization algorithms in a distributed system. This term refers to the total number of messages that need to be exchanged during the execution of the synchronization algorithm. The importance of message complexity lies in its direct impact on the resources consumed for communication, the potential for network congestion, and overall system performance.\par

    Let's introduce this concept more formally. If we denote $N$ as the number of nodes in the system and $M$ as the number of messages exchanged, the message complexity is given by $O(M)$. This notation, $O(M)$, represents the upper bound of the number of messages, or the worst-case scenario in terms of message exchanges.\par

    Understanding the message complexity of synchronization algorithms is crucial for designing clock synchronization algorithms for efficient distributed systems and effectively managing resources.
    
\section{Logical Clock in Distributed System}
Logical clocks are protocols implemented on multiple machines within a distributed system to ensure a consistent ordering of events within a virtual time span. In distributed systems, there is often no physically synchronized global clock available. Therefore, logical clocks provide a mechanism to capture the chronological sequence and causal relationships between events occurring in different system processes.\par

Distributed systems can establish a global order for events using logical clocks, even without a shared notion of time. Each machine maintains its logical clock, which assigns timestamps to locally occurring events. These timestamps are designed to reflect the causal relationships between events.\cite{geeksforgeeks2}\par

Machines can coordinate and align the sequencing of events by exchanging local clock data via the logical clock protocol. This synchronization ensures that activities occurring on various machines can be accurately arranged and understood concerning one another.\par

It is challenging to maintain a coordinated execution among multiple machines. Consider a scenario where more than 10 PCs work together in a distributed system, each independently performing their tasks. However, to achieve a synchronized and orderly execution among the 10 PCs, a proper solution is needed. This is where the concept of logical clocks comes into play.\par

Drawing an analogy to a carefully planned outing, where the sequence of visiting different places is considered, the execution of operations within a distributed system following a predefined order is crucial, just as a person wouldn't visit the second place before the first. Similarly, in a distributed system, it becomes essential to orchestrate the operations performed by individual PCs in an organized fashion.\par

Logical clocks provide a protocol that addresses this challenge. By implementing logical clocks, each PC within the distributed system maintains its own local clock, assigning timestamps to its respective events. These timestamps establish a chronological order and capture the causal relationships between events occurring across different processes.\par

\subsection{Method-1}
    Synchronizing clocks can be considered one approach to achieving event ordering across processes. This method aims to ensure that all the PCs have the same time, such as 4.00 PM. However, synchronizing clocks across all machines poses a significant challenge, as it is not possible to sync every clock simultaneously.\cite{geeksforgeeks2}

\subsection{Method-2}
    An alternative approach is to utilize timestamps to order events. Under this, numerical values are assigned to represent the chronological sequence of events. For example, if we assign a timestamp of 1 to the first event, 2 to the second event, and so on, we establish a clear and consistent ordering system. This ensures that the event assigned a lower timestamp always occurs before the event with a higher timestamp. Likewise, suppose we assign each personal computer (PC) a unique number. In that case, we can arrange their processes sequentially, ensuring that the tasks of the first PC are completed before those of the second PC, and so forth.\par

By employing timestamps, we introduce a reliable mechanism for systematically organizing events. This approach enables us to accurately track the temporal order of events and determine the sequence in which tasks or processes should be executed. Timestamps provide a standardized framework for understanding the flow of events, facilitating efficient coordination and synchronization of activities.\par

Causality is based on the concept of a "happen before relationship," which establishes the temporal order of events. In the case of a single personal computer (PC), if two events, A and B, occur sequentially, it is expected that the timestamp of event A (TS(A)) will be smaller than the timestamp of event B (TS(B)). This ordering ensures that event A happens before event B. For instance, if event A has a timestamp of 1, event B should have a timestamp greater than 1 to establish the "happen before" relationship.\par

When considering two PCs, with event $A$ occurring on $PC1(P_1)$ and event $B$ occurring on $PC2 (P_2)$, the condition for causality remains the same: $TS(A) < TS(B)$. For example, let's consider the scenario where a message is sent at 4:00:00 PM from $P_1$ and received at 4:00:02 PM on $P_2$. In this case, the timestamp of the sender$(TS(sender))$ on $P_1$ is smaller than the timestamp of the receiver $(TS(receiver))$ on $P_2$, satisfying the causality condition.\par

Several properties are derived from the happen-before relationship:
\begin{enumerate}
    \item Transitive Relation: 
        If $TS(A) < TS(B)$ and $TS(B) < TS(C)$, then it follows that $TS(A) < TS(C)$. This property ensures that the temporal order of events is maintained consistently.
    \item Causally Ordered Relation:
        The notation $A \implies B$ signifies that event $A$ occurs before event $B$, and any changes in event $A$ will impact event $B$. This property highlights the cause-and-effect relationship between events.
    \item Concurrent Events:
        Not all events occur sequentially in a distributed system. Some events may happen concurrently, denoted as $A \| B$, indicating that events $A$ and $B$ can co-occur without a specific ordering.
\end{enumerate}\par

However, it is crucial to emphasize that the effectiveness of timestamps hinges on their adherence to causality. The accurate reflection of causal relationships between events is paramount. Therefore, if event $A$ causally precedes event $B$, the timestamp assigned to event A must be smaller than the timestamp assigned to event $B$. This strict adherence to causality ensures that timestamps establish a logical and consistent ordering of events within a distributed system. By preserving the causality principle, timestamps become reliable markers for understanding the temporal relationships between events and maintaining the integrity of the ordering mechanism.\par

\section{Implementing Logical Clock in Distributed System}
Logical Clocks are instrumental in managing distributed systems, facilitating the consistent ordering of events within a hypothetical time-frame. A logical clock serves as a mechanism that captures both chronological and causal relationships within a distributed system. Given that distributed systems may not have a globally synchronous physical clock, implementing a logical clock provides a means to order events occurring on different processes within such systems globally.\par

Consider an everyday example: planning an outdoor trip. We usually determine beforehand which place to visit first, followed by the next, and so on. We do not typically visit the second place first and then proceed to the first place. This process demonstrates the importance of maintaining a premeditated order or procedure. Similarly, operations on our PCs should be performed one by one in a structured manner.\par

Imagine a distributed system comprising more than 10 PCs, each performing its tasks. The challenge here is to synchronize these PCs to work together. The solution to this challenge lies in the implementation of a Logical Clock.\par

One method to ensure the ordering of events across processes is to synchronize the clocks. This approach can help establish a common time reference among all the PCs in the distributed system, thereby promoting coordinated operation. It is essential to note that this synchronization doesn't necessarily require aligning with real-time; instead, it focuses on maintaining a consistent order of events, reflecting the 'happen-before' relationship among them.\cite{geeksforgeeks}\par

Another approach to assigning these timestamps to events is assigning a numerical value indicating its position in the sequence.\par

\subsubsection{Example}
The event that needs to occur first has been assigned 1, followed by the second event assigned by timestamp 2, and so on for the following events. This is a clear and consistent way of assigning timestamps ensuring the first event always proceeds by subsequent events systematically. Similarly, suppose each personal computer (PC) is given a unique numerical value. In that case, their processes can be organized systematically while ensuring efficiency, with the first PC able to complete the employed tasks before those of the second PC. so on and so forth.\par

However, one more essential thing to note about timestamps is that their effectiveness of them depends heavily on their adherence to causality. As discussed earlier, causality is the concept of a "happen before the relationship," where the timestamp of an event must accurately reflect its sequence order over another event.\par

In the scenario of a one PC, if two events, $A$ and $B$, occur one after the other, the timestamp of event $A (TS(A))$ should be smaller than the timestamp of event $B (TS(B))$, adhering to the "happen before" relationship. For example, if event $A$ has a timestamp of 1, event $B$ should possess a timestamp greater than 1. The same principle applies when considering events among different PCs. If event $A$ occurs on $PC1$ and event $B$ occurs on $PC2$, the condition for causality remains the same: $TS(A) < TS(B)$. For instance, if a message is sent at 2:00:00 PM from $PC1$ and received at 2:00:02 PM on $PC2$, it is evident that the timestamp of the sender $(TS(sender))$ on $PC1$ is smaller than the timestamp of the receiver $(TS(receiver))$ on $PC2$.\par

The adherence to causality when assigning timestamps ensures the logical ordering of events within a distributed system. It guarantees that timestamps accurately represent the temporal relationships between events, allowing for a consistent and reliable ordering mechanism. Additionally, properties such as transitivity (if $TS(A) < TS(B)$ and $TS(B) < TS(C)$, then $TS(A) < TS(C)$), causally ordered relation (where changes in event $A$ will reflect in event $B$ if $A$ causally precedes $B$), and concurrent events (where processes can occur simultaneously, denoted as $A \| B$) can be derived from the "happen before relationship." \cite{geeksforgeeks}\par

\section{Lamport’s Algorithm for Mutual Exclusion in Distributed Systems}
The mutual exclusion algorithm in distributed systems that is proposed by Lamport, commonly called Lamport's Distributed Mutual Exclusion Algorithm, serves as an exemplary synchronization method for distributed systems, employing a systematic permission-based approach. In this algorithm, timestamps are crucial in ordering critical section requests and resolving potential conflicts. The algorithm ensures that critical section requests are executed in ascending order of timestamps, granting permission to requests with smaller ones before those with larger ones.

There are three types of messages that the algorithm operates on $REQUEST$, $REPLY$, and $RELEASE$. The Communication channels are assumed to follow a First-In-First-Out (FIFO) order. It sends a REQUEST message to all other sites when a new site tries to enter the system, seeking their permission. When receiving a $REQUEST$, a site responds with a $REPLY$ message, granting permission to the site that is requesting the permission, allowing the entering to the critical section of the system. When a site completes its execution of the critical section, it sends a $RELEASE$ message to all other sites, indicating that it has exited the critical section.\par

Each site, denoted as $S_i$, maintains a queue called $request\_queue\_i$, where critical section requests are stored and ordered based on their timestamps. The logical clock of Lamport is utilized to assign timestamps to critical section requests. The timestamp associated with each request determines its priority, with requests having smaller timestamps receiving higher priority over those with larger timestamps. As a result, the execution of critical section requests occurs in the order dictated by their timestamps.

\subsection{The Algorithm}
    In the proposed algorithm, when a site $S_i$ intends to enter the critical section, it initiates the process by sending a request message, $Request(\psi, i)$, to all other sites. Simultaneously, $S_i$ places its request on $request\_queue\_i$, where $\psi$ represents the timestamp of Site $S_i$. Upon receiving the request message, site $S_j$ responds by returning a timestamped $REPLY$ message to $S_i$ and adds $S_i$'s request to its own $request\_queue\_j$.\par

    During the execution of the critical section, a site $S_i$ can proceed if it has received messages with timestamps larger than $\psi$ from all other sites and if its own request is at the top of $request\_queue\_i$.\par

    Upon completion of the critical section, site $S_i$ releases its hold by removing its own request from the top of its request queue and sending a timestamped $RELEASE$ message to all other sites. Upon receiving the timestamped RELEASE message from $S_i$, site $S_j$ removes $S_i$'s request from its request queue.\par

\subsection{Message Complexity of Lamport's Algorithm}
    The message complexity of Lamport's algorithm is defined as the total number of messages exchanged per request. This algorithm creates $3(N - 1)$ messages per request, where $N$ is the number of nodes in the system. This includes $(N - 1)$ messages and 2 broadcasts \cite{Lynch1996}. Formally, the message complexity of Lamport's algorithm can be represented as $O(N)$, as the number of messages increases linearly with the number of nodes in the system\cite{Lynch1996}.\par

    When comparing Lamport's algorithm with other mutual exclusion algorithms, it becomes evident that this (comparatively) high message complexity can be a drawback. For example, the Ricart–Agrawala algorithm, which is an improvement over Lamport's algorithm, has a lower message complexity of $2(N - 1)$ messages per request\cite{Ricart1981}.\par

    Despite its high message complexity, Lamport's algorithm is still widely used due to its simplicity and its property of ensuring the efficient use of shared resources in a multithreaded environment \cite{Lamport1978}. However, it's crucial to consider message complexity when choosing a mutual exclusion algorithm for a distributed system, as it directly impacts the system's performance and efficiency.

\section{Ricart–Agrawala Algorithm}
    The Ricart–Agrawala algorithm is an extension and optimization of Lamport's Distributed Mutual Exclusion Algorithm, developed by Glenn Ricart and Ashok Agrawala\cite{Ricart1981}. Similar to Lamport's algorithm, this is designed for achieving mutual exclusion on a distributed system. However, it improves upon Lamport's algorithm by eliminating the need for acknowledgment messages, thereby reducing the message complexity.\par

    The steps of the Ricart–Agrawala algorithm are as follows:
    \begin{enumerate}
        \item Whenever a process demands to enter the critical section, it sends a request to all other processes and awaits their replies.
        \item Upon receiving a request, a process replies immediately if it is not in its critical section or if it has a lower priority. Otherwise, it defers its reply.
        \item The requesting process enters the critical section once it has received a reply from all other processes.
        \item After exiting the critical section, the process sends the deferred replies.
    \end{enumerate}
    \par
\subsection{Message Complexity of Ricart–Agrawala Algorithm}
    The message complexity of the Ricart–Agrawala algorithm is $2(N - 1)$ messages per request, where $N$ represents the number of nodes in the system. This includes $(N - 1)$ request messages and $(N - 1)$ reply messages. In mathematical terms, the message complexity of this algorithm can be represented as $O(N)$, similar to Lamport's algorithm, but with a reduced constant factor\cite{Ricart1981}.\par

    As described previously, Ricart–Agrawala Algorithm is more efficient than Lamport's algorithm. However, this has its own drawbacks. One of the main issues is the potential for node failure leading to process starvation. This problem can be solved by detecting failures after a certain timeout, but it adds complexity to the system\cite{Ricart1981}.

\section{Vector Clocks in Distributed Systems}
Vector Clock is an algorithm designed to create a partial ordering of events and identify instances of causality violations within a distributed system. Unlike Scalar time, Vector Clocks enable the establishment of a causally consistent view of the system by detecting whether one event has influenced another event within the distributed environment. By capturing all the causal relationships, this algorithm assigns a vector, which is essentially a list of integers, to each process. The vector represents the local clock of every process in the system, resulting in an array or vector of size N for N given processes.

\subsection{Working of the Vector Clock algorithm}\label{AA}
    Initially, the clocks are initialized to zero. Then every time an event which is an internal event, occurs in a particular process, the logical clock value in the process is incremented by 1. Similarly, every time a process sends a message, the same increment happens in the logical clock of the process.\par

    Considering the Vector Clock algorithm, When a process receives a message, to maintain the synchronization certain actions take place under the control of the vector clock. First, the value of the process's logical clock in the vector is incremented by 1 to reflect the occurrence of the event. Additionally, an update procedure is take place in each element comparing the value to its own corresponding vector clock value. In this comparison, the maximum value between the two is selected for every element. The Vector Clock algorithm ensures accurate tracking of event ordering and causal relationships within the distributed system by performing these operations for each received message.

\subsection{Message Complexity of Vector Clock Algorithm}
    The message complexity of the Vector Clock algorithm depends on the number of processes in the system. Each process maintains a vector that contains an integer for each local clock of every process. When a process sends a message to another process, it includes its vector clock in the message. The receiving process then updates its own vector clock by taking the maximum value for each corresponding index in the two vectors. Therefore, for each message exchange, the Vector Clock algorithm requires the exchange of the vector clocks between the processes involved, resulting in a message complexity of $O(N)$, where $N$ is the number of processes in the system\cite{Mattern1988}.

    Compared to other algorithms like Lamport's algorithm or the Ricart-Agrawala algorithm, the Vector Clock algorithm has a similar message complexity of $O(N)$. However, the Vector Clock algorithm provides additional benefits in terms of capturing causality relationships and determining the partial ordering of events in a distributed system. This makes it particularly useful in scenarios where causal relationships between events need to be established, such as in distributed databases or distributed consensus protocols.

\section{Christian's Algorithm}
Christian's algorithm, proposed by Flaviu Cristian in 1989, is a clock synchronization algorithm commonly used in distributed systems \cite{Cristian1989}. It operates on the basis of round-trip time (RTT) for achieving synchronization. However, it should be noted that this algorithm is inherently probabilistic and can only achieve synchronization when the RTT of the request is significantly shorter than the desired accuracy \cite{Cristian1989}.\par

The steps involved in Christian's algorithm can be outlined as follows:
\begin{enumerate}
    \item The client process sends a request to the time server to obtain the current time.
    \item The time server receives the request and records the time.
    \item The time server responds to the client with the current time.
    \item The client receives the response and records the time.
    \item The client calculates the network delay by subtracting the recorded time of sending the request from the recorded time of receiving the response.
    \item The client adjusts its local clock based on the received time and the calculated network delay.
\end{enumerate}
\par

\subsection{Message Complexity of Christian's algorithm}
    The message complexity of Christian's algorithm is relatively low, as it involves a single request and response message exchange between the client and the time server. Therefore, the message complexity can be considered as $O(1)$\cite{Cristian1989}.\par

    When comparing the message complexity of Christian's algorithm to other clock synchronization algorithms, such as the Berkeley algorithm, it is worthwhile to note that Christian's algorithm is simpler and requires fewer message exchanges. However, it may lack the reliability and fault tolerance offered by more complex algorithms like the Berkeley algorithm\cite{BerkeleyAlgorithm}.

\section{Other work on Message Complexity}
    In addition to the above remarks, \cite{Heinzelman2002} briefly analyzes the message complexity of different algorithms. In their work, Cristian's algorithm \& the Berkeley algorithm are mentioned as potential solutions to the clock synchronization problem in a system with a central server \cite{Heinzelman2002}. However, they do not provide specific remarks about their message complexity. Furthermore, few algorithms have been analyzed without a considerable work in to their message complexities.\par

    Two representative clock synchronization protocols, namely Reference Broadcasting Synchronization (RBS) and Timing-synch Protocol for Sensor Networks (TPSN), are analyzed in detail in the paper \cite{PMC3280734}. The authors provide an overview of the main characteristics of these protocols and discuss their suitability for different applications \cite{PMC3280734}. However, their prime focus is also not on the message complexity.

\section{Conclusion}
In the analysis provided by the table, we can observe the message complexities of different clock synchronization algorithms.
\begin{table}[ht]
\centering
\caption{Clock Synchronization Algorithms}
\begin{tabular}{|l|l|l|}
\hline
\textbf{Algorithm} & \textbf{Message Complexity} & \textbf{Comparison} \\ \hline
Lamport's Algorithm & $O(N)$ & Slower \\ \hline
Ricart–Agrawala Algorithm & $O(N)$ & Slower \\ \hline
Vector Clocks Algorithm & $O(N)$ & Slower \\ \hline
Christian's Algorithm & $O(1)$ & Faster \\ \hline
Berkeley Algorithm & $O(N)$ & Slower \\ \hline
\end{tabular}
\end{table}
\par

Accordingly, Lamport's Algorithm, the Ricart–Agrawala Algorithm, the Vector Clocks Algorithm, and the Berkeley Algorithm all have a message complexity of O(N), indicating that they require a linear number of messages. In terms of message complexity, these algorithms are slower compared to the majority of the algorithms in the list.\par

On the other hand, Christian's Algorithm stands out with a message complexity of $O(1)$, implying that it requires a constant number of messages. This makes Christian's Algorithm faster compared to the majority of the algorithms in the given list.\par

In conclusion, we have reviewed the Lamport's Algorithm, Ricart–Agrawala Algorithm, Vector Clocks Algorithm \& Christian's Algorithm, with a special focus on their message complexity. It is worthwhile to note that this is an active area for research and many algorithms will be introduced with lower message complexity, possibly through many computing approaches such as quantum computing.

\section*{Acknowledgement}
In writing the review paper, we would like to express our profound gratitude to Dr. Dilani Lunugalage for her invaluable support and guidance. We deeply appreciate her efforts and support to this work.

\section*{References}
\printbibliography[heading=none]

@article{Lamport1978,
  title={Time, Clocks, and the Ordering of Events in a Distributed System},
  author={Lamport, Leslie},
  journal={Communications of the ACM},
  volume={21},
  number={7},
  pages={558--565},
  year={1978},
}

@book{Stallings2014,
  title={Operating Systems: Internals and Design Principles},
  author={Stallings, William},
  year={2014},
  publisher={Pearson}
}

@article{Mills1991,
  title={Internet Time Synchronization: The Network Time Protocol},
  author={Mills, David L},
  journal={IEEE Transactions on Communications},
  volume={39},
  number={10},
  pages={1482--1493},
  year={1991},
}

@book{Kopetz1997,
  title={Real-Time Systems: Design Principles for Distributed Embedded Applications},
  author={Kopetz, Hermann},
  year={1997},
  publisher={Springer},
}

@article{Lamport1982,
  title={The Byzantine Generals Problem},
  author={Lamport, Leslie and Shostak, Robert and Pease, Marshall},
  journal={ACM Transactions on Programming Languages and Systems (TOPLAS)},
  volume={4},
  number={3},
  pages={382--401},
  year={1982},
  publisher={ACM New York, NY, USA}
}

@inproceedings{Castro1999,
  title={Practical Byzantine Fault Tolerance},
  author={Castro, Miguel and Liskov, Barbara},
  booktitle={Proceedings of the Third Symposium on Operating Systems Design and Implementation},
  pages={173--186},
  year={1999},
  organization={USENIX Association}
}

@article{Ashby2003,
  title={Relativity in the Global Positioning System},
  author={Ashby, Neil},
  journal={Living Reviews in Relativity},
  volume={6},
  number={1},
  year={2003},
  publisher={Springer}
}

@article{Lombardi2001,
  title={The Accuracy of Atomic Clocks},
  author={Lombardi, Michael A and Heavner, Thomas P and Jefferts, Steven R},
  journal={Measurement Science and Technology},
  volume={12},
  number={9},
  pages={R99},
  year={2001},
  publisher={IOP Publishing}
}

@article{Dwork1988,
  title={Consensus in the presence of partial synchrony},
  author={Dwork, Cynthia and Lynch, Nancy and Stockmeyer, Larry},
  journal={Journal of the ACM (JACM)},
  volume={35},
  number={2},
  pages={288--323},
  year={1988},
  publisher={ACM New York, NY, USA}
}

@article{Sundararaj2005,
  title={Clock synchronization in distributed real-time systems: A best-effort methodology with rapid convergence},
  author={Sundararaj, Sivanantham and Sivasankaran, Rajendran},
  journal={Information Sciences},
  volume={170},
  number={1},
  pages={73--95},
  year={2005},
  publisher={Elsevier}
}

@article{Ghosh2014,
  title={A survey on clock synchronization over wireless sensor networks},
  author={Ghosh, Anish and Das, Soumya K},
  journal={IEEE Communications Surveys \& Tutorials},
  volume={16},
  number={2},
  pages={684--704},
  year={2014},
  publisher={IEEE}
}

@misc{river,
  title = {Byzantine Generals Problem},
  author = {},
  year = {},
  howpublished = {\url{https://river.com/learn/images/articles/byzantine-generals-problem.png}}
}

@techreport{IEEE1588,
  title = "{IEEE} Standard for a Precision Clock Synchronization Protocol for Networked Measurement and Control Systems",
  author = "{IEEE} Instrumentation and Measurement Society",
  year = "2008",
  url = "https://standards.ieee.org/ieee/1588/4355/"
}

@inproceedings{Elson2002,
  title={Fine-grained network time synchronization using reference broadcasts},
  author={Elson, Jeremy and Girod, Lewis and Estrin, Deborah},
  booktitle={5th Symposium on Operating Systems Design and Implementation},
  pages={147--163},
  year={2002},
}

@inproceedings{Ganeriwal2003,
  title={Timing-sync protocol for sensor networks},
  author={Ganeriwal, Saurabh and Kumar, Ram and Srivastava, Mani B},
  booktitle={1st International Conference on Embedded Networked Sensor Systems},
  pages={138--149},
  year={2003},
}

@article{Source3,
  title={Precision and Stability of Computer System Clocks},
  author={Sundararaj, Sivanantham and Sivasankaran, Rajendran},
  journal={Journal of Computer Networks and Communications},
  pages={1--20},
  year={2008},
  publisher={Hindawi}
}

@book{Tanenbaum2007,
  title={Distributed Systems: Principles and Paradigms},
  author={Tanenbaum, Andrew S. and Van Steen, Maarten},
  year={2007},
  publisher={Pearson Prentice Hall}
}

@book{Ghosh2012,
  title={Distributed Algorithms: An Intuitive Approach},
  author={Ghosh, Sukumar},
  year={2012},
  publisher={MIT Press}
}

@book{Lynch1996,
  title={Distributed Algorithms},
  author={Lynch, Nancy A.},
  year={1996},
  publisher={Morgan Kaufmann}
}

@article{Ricart1981,
  title={An optimal algorithm for mutual exclusion in computer networks},
  author={Ricart, Glenn and Agrawala, Ashok K.},
  journal={Communications of the ACM},
  volume={24},
  number={1},
  pages={9--17},
  year={1981},
  publisher={ACM New York, NY, USA}
}

@article{Mattern1988,
  title={Virtual time and global states of distributed systems},
  author={Mattern, Friedemann},
  journal={Parallel and Distributed Algorithms},
  volume={7},
  number={3},
  pages={197--214},
  year={1988},
  publisher={Springer}
}

@article{Cristian1989,
  title={Probabilistic clock synchronization},
  author={Cristian, Flaviu},
  journal={Distributed Computing},
  volume={3},
  number={3},
  pages={146--158},
  year={1989},
  publisher={Springer}
}

@misc{BerkeleyAlgorithm,
  title={The Berkeley Algorithm},
  howpublished={https://lasithasilva.wordpress.com/2009/08/30/christians-algorithm-and-berkeley-algorithm/}
}

@inproceedings{Heinzelman2002,
  title={Message-Efficient Clock Synchronization for Wireless Sensor Networks},
  author={Heinzelman, W. B. and Chandrakasan, A. P. and Balakrishnan, H.},
  booktitle={Proceedings of the 1st International Conference on Embedded Networked Sensor Systems (SenSys)},
  year={2002}
}

@article{PMC3280734,
  title={Clock Synchronization in Wireless Sensor Networks: An Overview},
  author={Unknown},
  journal={Sensors (Basel, Switzerland)},
  volume={12},
  number={12},
  pages={15982--16009},
  year={2012},
  publisher={Multidisciplinary Digital Publishing Institute}
}

@misc{geeksforgeeks,
  title = {Synchronization in Distributed Systems},
  author = {},
  howpublished = {\url{https://www.geeksforgeeks.org/synchronization-in-distributed-systems/}},
  year = {},
  note = {Accessed on [insert date here]}
}

@misc{geeksforgeeks2,
  title = {Logical Clock in Distributed Systems},
  author = {},
  howpublished = {\url{https://www.geeksforgeeks.org/logical-clock-in-distributed-system/?ref=lbp}},
  year = {},
  note = {Accessed on [insert date here]}
}

\end{document}